\documentclass[11pt,a4paper]{article}

\usepackage[utf8]{inputenc}
\usepackage[T1]{fontenc}
\usepackage{mathptmx}
\usepackage{microtype}
\usepackage[margin=1in]{geometry}
\usepackage{setspace}
\usepackage{parskip}
\usepackage{amsmath,amssymb}
\usepackage{booktabs}
\usepackage{array}
\usepackage{multirow}
\usepackage{makecell}
\usepackage{tabularx}
\usepackage{longtable}

\usepackage{graphicx}
\usepackage{rotating}
\usepackage[table]{xcolor}
\definecolor{accent}{HTML}{1B4A73}
\definecolor{pos}{HTML}{1B7837}
\definecolor{neg}{HTML}{B35806}
\definecolor{softbg}{HTML}{F7F6F1}
\definecolor{ruleg}{HTML}{E5E3DD}

\usepackage{float}
\usepackage{caption}
\usepackage{listings}
\usepackage[hidelinks,breaklinks=true]{hyperref}
\hypersetup{
  colorlinks=true,
  linkcolor=accent,
  citecolor=accent,
  urlcolor=accent,
  pdftitle={Towards Nexus-Score: Metadata Gaps Limit Scholarly AI Attribution},
  pdfauthor={Aadi Narayana Varma Dantuluri, Sushrut Thorat, Paras Chopra},
}

\usepackage{titlesec}
\titleformat{\section}{\normalfont\large\bfseries\color{accent}}{\thesection}{0.6em}{}
\titleformat{\subsection}{\normalfont\normalsize\bfseries}{\thesubsection}{0.6em}{}
\titleformat{\subsubsection}{\normalfont\normalsize\itshape}{\thesubsubsection}{0.6em}{}
\titlespacing*{\section}{0pt}{1.4em}{0.4em}
\titlespacing*{\subsection}{0pt}{1.0em}{0.3em}

\usepackage{tikz}
\usepackage{tcolorbox}
\tcbuselibrary{skins,breakable}
\newtcolorbox{abstractbox}{
  enhanced, breakable, colback=softbg, colframe=accent,
  boxrule=0pt, leftrule=2pt, arc=1pt,
  left=10pt, right=10pt, top=8pt, bottom=8pt,
  fontupper=\small,
}
\newtcolorbox{caveatbox}{
  enhanced, breakable, colback=yellow!8, colframe=yellow!50!brown,
  boxrule=0pt, leftrule=2pt, arc=1pt,
  left=8pt, right=8pt, top=6pt, bottom=6pt,
  fontupper=\small,
}
\newtcolorbox{nexusbox}{
  enhanced, breakable, colback=blue!3, colframe=accent,
  boxrule=0.5pt, arc=1pt,
  left=8pt, right=8pt, top=5pt, bottom=5pt,
  before skip=5pt, after skip=5pt,
  fontupper=\normalsize,
}
\newtcolorbox{placeholderbox}{
  enhanced, breakable, colback=yellow!28, colframe=yellow!70!black,
  boxrule=0.4pt, arc=1pt,
  left=8pt, right=8pt, top=6pt, bottom=6pt,
  fontupper=\small,
}

\newcommand{\code}[1]{\texttt{\small #1}}
\newcommand{\nexus}{\textsc{Nexus-Score}}

\definecolor{orcidgreen}{HTML}{A6CE39}
\newcommand{\orcidicon}{%
  \begin{tikzpicture}[baseline=-0.68ex]
    \fill[orcidgreen] (0,0) circle (0.82ex);
    \node at (0,0) {\textcolor{white}{\fontsize{5.5}{5.5}\selectfont\bfseries iD}};
  \end{tikzpicture}%
}
\newcommand{\orcidlink}[1]{\href{https://orcid.org/#1}{\orcidicon}}
\title{\vspace{-2.6em}\includegraphics[height=1.1cm]{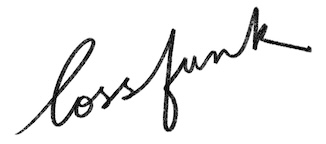}\\[1.0em]
\textbf{\texorpdfstring{Towards Nexus-Score:\\ Metadata Gaps Limit Scholarly AI Attribution}{Towards Nexus-Score: Metadata Gaps Limit Scholarly AI Attribution}}\vspace{0.4em}}
\author{
  Aadi~Narayana~Varma~Dantuluri~\orcidlink{0000-0001-8551-3638}\thanks{Correspondence: \href{mailto:aadi@nexus-score.org}{aadi@nexus-score.org}.}\\
  \small \emph{\href{https://nexus-score.org}{nexus-score.org}}\\[0.35em]
  Sushrut~Thorat~\orcidlink{0000-0003-2276-5621}\\
  \small \emph{Lossfunk, \href{https://lossfunk.com}{lossfunk.com}}\\[0.35em]
  Paras~Chopra~\orcidlink{0009-0007-9982-0015}\\
  \small \emph{Lossfunk, \href{https://lossfunk.com}{lossfunk.com}}\\
}
\date{}

\begin{document}
\maketitle
\vspace{-1.8em}

\begin{abstractbox}
\noindent\textbf{Abstract.}
Artificial intelligence systems increasingly mediate how science is found and credited. We asked whether missing metadata prevents AI systems from crediting work. As a boundary test, an AI system citing without access to task-relevant paper lists often produced out-of-list identifiers, some fabricated. We then tested the mechanism in real scholarly infrastructure by using OpenAlex records to hide or restore author, institution, funder, reference, and text-access links while holding works and tasks fixed. Restoring the relevant link made the corresponding attribution possible; restoring the wrong kind did not, with 0 correct answers across 469 completed mismatched tests. Thus, in these tasks, one metadata facet did not substitute for another. Missing links led to invented answers, refusals, or tool-budget exhaustion, and web search did not recover hidden author links. In sum, AI systems credited work only when record connections were visible or recoverable. This motivates Nexus-Score, a record-level check for metadata gaps, to guide repair and help prepare the scholarly record for AI-mediated use.
\end{abstractbox}


% =====================================================================
\section{Scholarly records as AI substrate}
\label{sec:intro}

Scientific papers are written for people. Scholarly databases turn those papers into machine-readable descriptions: title, digital object identifier (DOI), authors, affiliations, funders, reference lists, licenses, and links to text. We call one such description a \emph{record}. OpenAlex is one large example of this kind of record system \cite{openalex2022}. Independent audits of OpenAlex and other open scholarly databases show that metadata coverage is uneven across fields, sources, and metadata types \cite{delgadoquiros2024completeness,culbert2025openalex,cespedes2025linguistic}. FAIR data principles make the same general point that research objects become more useful when they are findable, accessible, interoperable, and reusable \cite{wilkinson2016fair,wilkinson2018metrics}. Open-science and open-research-information efforts make a related infrastructure claim: research is easier to check and reuse when its records, evidence, and governance are transparent \cite{barcelona,nosek2015top,munafo2017manifesto}.

When a scholarly AI system queries a database or retrieval tool, these records are what the system receives. If the record exposes that a work was written by a given author, funded by a given funder, cites a given paper, or has a full-text route, the system can use that connection. If the record does not expose the connection, the system faces a different problem. It may guess from memory, search elsewhere, substitute a nearby record, or decline to answer.

Citation hallucination is therefore not only a model problem, even though model behavior clearly matters \cite{walters2023citations,chelli2024hallucination,alkaissi2023chatgpt,ji2023survey}. For scholarly attribution, there is also a record problem. A system asked to name a funder needs a work-funder link. A system asked to trace a citation needs a reference list. A system asked to read a study needs an access route to the text. These are not decorative metadata fields. They are the evidence that a tool-using system can check.

The experiments below need names for these checkable links. We use \emph{connection} or \emph{edge} for an explicit relation in a record, such as ``this author wrote this work'' or ``this work cites that work.'' We group related connections into five \emph{facets}: Provenance, People, Organizations, Funding, and Access. Provenance covers where the record came from and what the work cites. People covers authors. Organizations covers affiliations. Funding covers funders and awards. Access covers whether the text can be reached. We use \emph{agent} for a language model that calls tools to query an external scholarly database, rather than answering from memory alone.

If these links matter, then scholarly records need a way to be checked before they are used by artificial intelligence systems. \nexus\ is proposed as such a verifier. It does not ask whether a paper is good, important, prestigious, or highly cited. A low-scoring paper may be scientifically important; a high-scoring paper may be trivial. The narrower question is whether the record exposes the connections needed for attribution, reading, and reuse. A public publisher-level implementation applies this framing to Crossref metadata coverage; the experiments here instead use a record-level OpenAlex prototype (Section~\ref{sec:nexus}) \cite{dantuluri2026nexusscore}. This paper tests the need for that kind of verifier. We first ask whether a model cites differently with and without a retrieval record. We then hide and restore specific facets to test whether the corresponding attribution task fails and recovers. Finally, we ask whether naturally sparse records and inference-time rescue strategies show the same problem. The results motivate \nexus\ as a record verifier and point to metadata repair as a downstream infrastructure need. The supplementary materials give the model settings, tool interface, record views, task definitions, outcome rules, scoring details, and reproducibility information needed to audit each step.

% =====================================================================
\section{Retrieval constrains citation errors}
\label{sec:norag}

We first tested the boundary case in which the model has no scholarly substrate at all. The task asked the same model to answer handwritten questions about p53 and apoptosis, with each answer required to cite papers by DOI. In the retrieval-augmented condition, the model received a candidate list of real papers from a p53 corpus and was instructed to cite only from that list. In the no-retrieval condition, it had to cite from memory. Both conditions required DOIs, so a script could compare cited identifiers with the p53 corpus and check out-of-corpus identifiers at \code{doi.org} (Supplementary Methods, section~\ref{sec:appendix-norag}).

With a retrieved candidate set, every cited DOI was drawn from the corpus and no response contained a fabricated DOI (Supplementary Table~\ref{tab:norag}). Without retrieval, none of the 65 unique cited DOIs was in the corpus, and 15 of 20 calls contained at least one fabricated DOI. Some of those out-of-corpus DOIs resolved at \code{doi.org}, so they were real identifiers, but they were not verified as answers to the task. Others were fabricated.

This contrast is not a balanced comparison: the no-retrieval arm was stopped early and covers fewer questions. We use it for a narrower purpose. It shows why an external scholarly record is needed before we can ask which parts of that record matter. The next experiment therefore keeps the work, model, task, and tool interface fixed while changing only which record connections the agent can see.

% =====================================================================
\section{Nexus-Score measures visible record links}
\label{sec:nexus}

If missing connections matter, a useful verifier should tell us which are present and which are absent. \nexus\ is our proposed way to do that. The experiments use it narrowly, as a candidate record verifier whose ingredients can be tested. The box below defines those ingredients and distinguishes the public publisher-level implementation from this paper's record-level prototype; the latter uses equal facet weights and prespecified nonuniform signal weights and is not a deployed universal standard (Supplementary Methods).

\begin{nexusbox}
\textbf{Nexus-Score at a glance.} \nexus\ checks whether selected scholarly-record connections are usable by downstream systems. It organizes them into five facets: \textbf{Provenance} (record source, identifiers, and references), \textbf{People} (authors and persistent identities), \textbf{Organizations} (affiliations and organization identifiers), \textbf{Funding} (funders and awards), and \textbf{Access} (licenses and routes to abstracts or full text). Within each facet, signals ask whether information is \emph{present, resolvable, specific, queryable,} and \emph{traceable}. The public implementation at \href{https://nexus-score.org}{nexus-score.org} summarizes Crossref coverage at publisher level using dimension weights of 25\%, 20\%, 15\%, 20\%, and 20\%, respectively \cite{dantuluri2026nexusscore}. This paper's OpenAlex prototype instead scores individual records with equal 20\% facet weights; within every facet, signal weights are 30\% presence, 25\% resolvability, 20\% specificity, 15\% queryability, and 10\% metadata provenance. In either version, a higher score indicates more usable metadata connections---not greater scientific quality, impact, or prestige. Project DOI: \href{https://doi.org/10.5281/zenodo.19217245}{10.5281/zenodo.19217245}.
\end{nexusbox}

The working corpus was deliberately mixed rather than uniformly rich. Across 625 works, the mean Nexus-Score composite was 0.65 (median 0.73), with 38\% of works below 0.5. Facet quality was uneven: Provenance was usually populated (mean 0.91), whereas Funding was often absent (mean 0.34, median 0.00; supplementary materials). This unevenness is the condition a verifier is meant to expose. The rest of the paper asks whether that exposed unevenness matters in practice. A record can have a DOI and still lack a funder, an institutional affiliation, a reference list, or a usable route to the text.

% =====================================================================
\section{Restored facets restore matching tasks}
\label{sec:diagonal}

\begin{sidewaysfigure}
\centering
\includegraphics[width=\textheight,height=\textwidth,keepaspectratio]{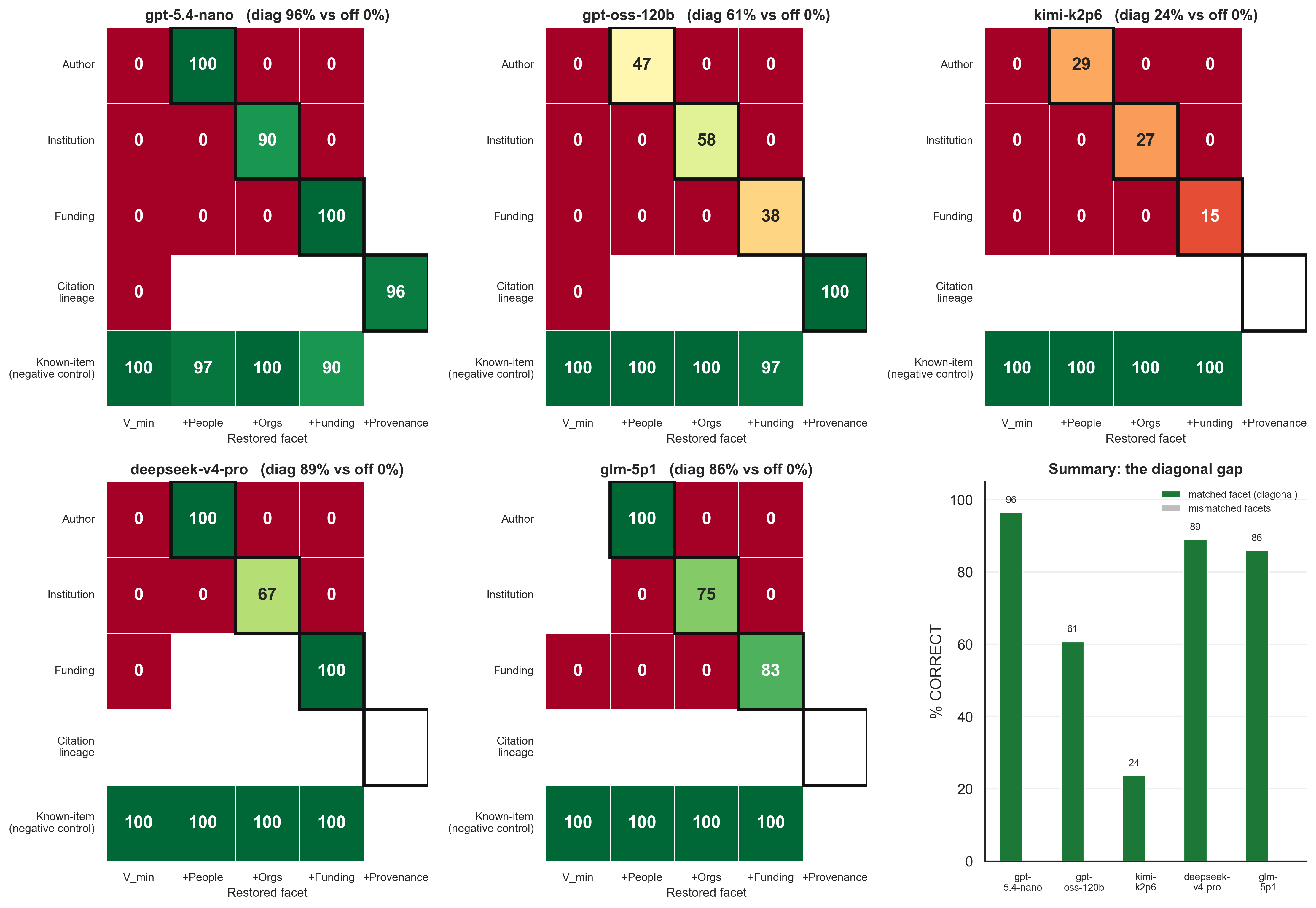}
\caption{\textbf{Restoring the missing metadata facet selectively restores the task that depends on it.} Each panel shows one model. Rows are tasks: naming an author, institution, funder, cited work, or the DOI of a known item. Columns are record views in which one facet is restored to an otherwise minimal record. The black outline marks the matched condition: the restored facet is the one the task requires. Accuracy is high in matched conditions and zero in completed mismatched conditions, showing that the facets expose specific connections rather than generic record richness. Across all five models, zero of 469 completed mismatched conditions was correct. Blank/white squares indicate unavailable or incomplete tests in that panel, not additional evidence for success or failure. The known-item row is a negative control: exact DOI lookup can succeed even in minimal views, and can worsen when irrelevant neighboring identifiers are shown.}
\label{fig:facet-diagonal}
\end{sidewaysfigure}

The core experiment asked whether the five metadata facets behave as distinct functional inputs. Starting from richly described records, we served the agent an otherwise minimal view and restored one facet at a time. The task suite then asked for the corresponding scholarly connection: author attribution, institution attribution, funding attribution, and citation lineage. If the facets matter, restoring People should help author attribution but not funder attribution; restoring Funding should help funder attribution but not author attribution. The expected result is therefore a diagonal pattern: the restored facet helps the matching task and not the others (Fig.~\ref{fig:facet-diagonal}).

That is what we observed. Among completed test conditions, matched restoration accuracy was 97\% for \code{gpt-5.4-nano}, 62\% for \code{gpt-oss-120b}, 22\% for \code{kimi-k2p6}, 92\% for \code{deepseek-v4-pro}, and 87\% for \code{glm-5p1}. By contrast, all completed mismatched conditions were 0\% correct. Across the five models, this means zero correct answers in 469 completed conditions where the restored facet did not match the task. The low Kimi value reflects many budget-exhausted runs rather than many wrong completed answers. Full denominators and intervals are reported in Supplementary Table~\ref{tab:diag}.

Citation lineage gave the same message for reference edges. In a two-model, 200-condition test, full records and Provenance-restored records supported citation lineage, whereas citation-masked and minimal records did not (Supplementary Table~\ref{tab:citation}). Thus the pattern is not limited to author or funder entities. It also applies to the reference links that underwrite scholarly synthesis.

This result supports facet-task dependence, not universal validation of the Nexus-Score composite. The experiment shows that the named facets correspond to separable attribution functions in this task suite. It does not establish that the current weights are optimal, that the composite predicts all downstream uses, or that the same values transfer unchanged to every scholarly graph.

% =====================================================================
\section{Missing links produce abstention or error}
\label{sec:failure}

When a needed connection is absent, an agent can fail in different ways. A weak or ungrounded system may invent an answer. A stronger grounded system may refuse or exhaust its tool budget. The second behavior is safer, but it is still not attribution. If the record lacks a funder connection and the agent says "I cannot determine the funder," the model has avoided hallucination; it has not credited the funder.

\begin{figure}[tbp]
\centering
\includegraphics[width=\linewidth]{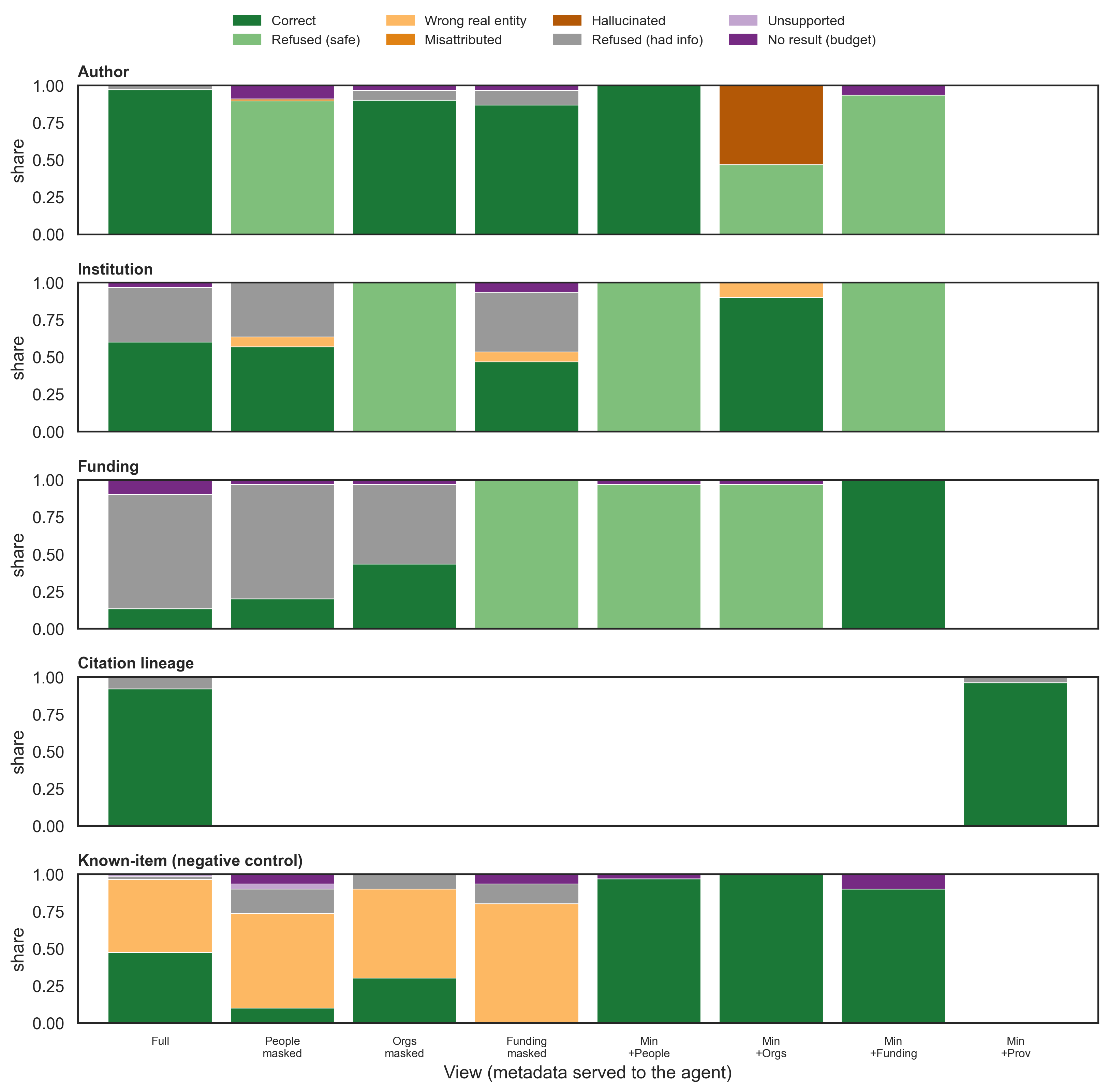}
\caption{\textbf{In the strongest grounded agent, missing connections usually produced abstention rather than invention.} Bars show outcomes classified by rule-based checks for \code{gpt-5.4-nano} across metadata views. Greens are substrate-respecting outcomes (correct or safe refusal), oranges and browns are invented or wrong edges, grey is refusal despite available information, and purple is budget exhaustion. Across completed rows excluding the literature-review probe, hallucinated or misattributed edges appeared in 18 of 1,211 responses (1.5\%, 95\% Wilson interval $[0.9, 2.3]$\%). This result describes this grounded agent, not all models. The bottom row is a control: for exact DOI lookup, rich records can distract the agent when the focal DOI is surrounded by references, supplement DOIs, datasets, and other true identifiers. The graph can be correct while the tool view is poorly matched to a narrow question.}
\label{fig:outcome}
\end{figure}

We therefore classified outcomes for the strongest grounded agent, \code{gpt-5.4-nano}, across the metadata views (Fig.~\ref{fig:outcome}). The categories separate correct answers, correct refusals, incorrect refusals, budget exhaustion, wrong real entities, misattributed edges, unsupported answers, and hallucinations. The classifier checks the relation, not just the names. A real author attached to the wrong paper is still wrong. Outcome definitions are given in Supplementary Table~\ref{tab:appendix-outcomes}.

In this grounded agent, unsafe completion was rare. Across completed rows excluding the literature-review probe, hallucinated or misattributed edges appeared in 18 of 1,211 responses (1.5\%, 95\% Wilson interval $[0.9, 2.3]$\%). The dominant failure modes were refusal and budget exhaustion. This result does not mean that missing metadata always makes AI systems refuse. The no-retrieval contrast shows that unsafe completion is possible, and the five-model restoration and rescue experiments show a wider range of non-completion: some models completed many cells, whereas others often exhausted their tool budgets. The safer claim is that missing substrate creates failure pressure whose form depends on the model and tools: invention in weakly grounded settings, refusal or budget exhaustion in stronger grounded settings, and non-attribution in all three.

The known-item control also prevents a simplistic reading of \nexus. More metadata is not the same as better prompting. If the question asks for the DOI of one named paper, a tool should return the focal record clearly. If it returns the target paper together with all references, related works, dataset links, supplement DOIs, and source metadata, the correct DOI may be buried among many true identifiers. \nexus\ measures whether the record contains useful connections; AI systems still need tools that expose the requested connection clearly.

% =====================================================================
\section{Sparse records limit testable attribution}
\label{sec:natural}

Controlled masking gives a clean counterfactual: the same work, model, and task, with one facet removed or restored. But the experiment would matter little if such missing-edge states never occurred in real records. We therefore asked whether low-scoring OpenAlex records naturally lacked the connections needed for the same attribution tasks.

We evaluated task buildability in a 25-work subset drawn from a 175-work low-scoring pool (Table~\ref{tab:natural}). Most attribution tasks could not be constructed because the OpenAlex record lacked the edge needed as ground truth.

\begin{table}[H]
\centering
\small
\begin{tabularx}{\linewidth}{lrlX}
\toprule
Task & \makecell[r]{Buildable tests} & Share & Reason for missingness \\
\midrule
Author attribution & 4 / 50 & 8\% & Most papers had no resolved author identifiers. \\
Institution attribution & 0 / 50 & 0\% & No institution identifiers in the pool. \\
Funding attribution & 0 / 50 & 0\% & No funder records. \\
Citation lineage & 0 / 50 & 0\% & No papers had \code{referenced\_works} populated. \\
Known-item DOI lookup & 50 / 50 & 100\% & Pool filter required a DOI. \\
\bottomrule
\end{tabularx}
\caption{\textbf{Natural low-scoring OpenAlex records often lacked the edge needed to define an attribution test.} The evaluated subset contains 25 works from a 175-work natural low-scoring pool. The dominant failure is task buildability, not model incapability: if the graph has no funder edge, no institution edge, or no reference list, the corresponding ground-truth question cannot be constructed from that substrate. The four buildable author-attribution tests were answered correctly by both tested models.}
\label{tab:natural}
\end{table}

These counts do not measure model performance. They do not show that the underlying relationships are absent in the world, nor that they could not be recovered from publisher pages, PDFs, or another graph. They show that the missing-edge state created experimentally is also a real state of the tested OpenAlex substrate. That matters for deployment, because an agent connected to that substrate cannot use edges that the substrate does not expose.

% =====================================================================
\section{Search does not reliably repair hidden links}
\label{sec:substitution}

\begin{figure}[tbp]
\centering
\includegraphics[width=\linewidth]{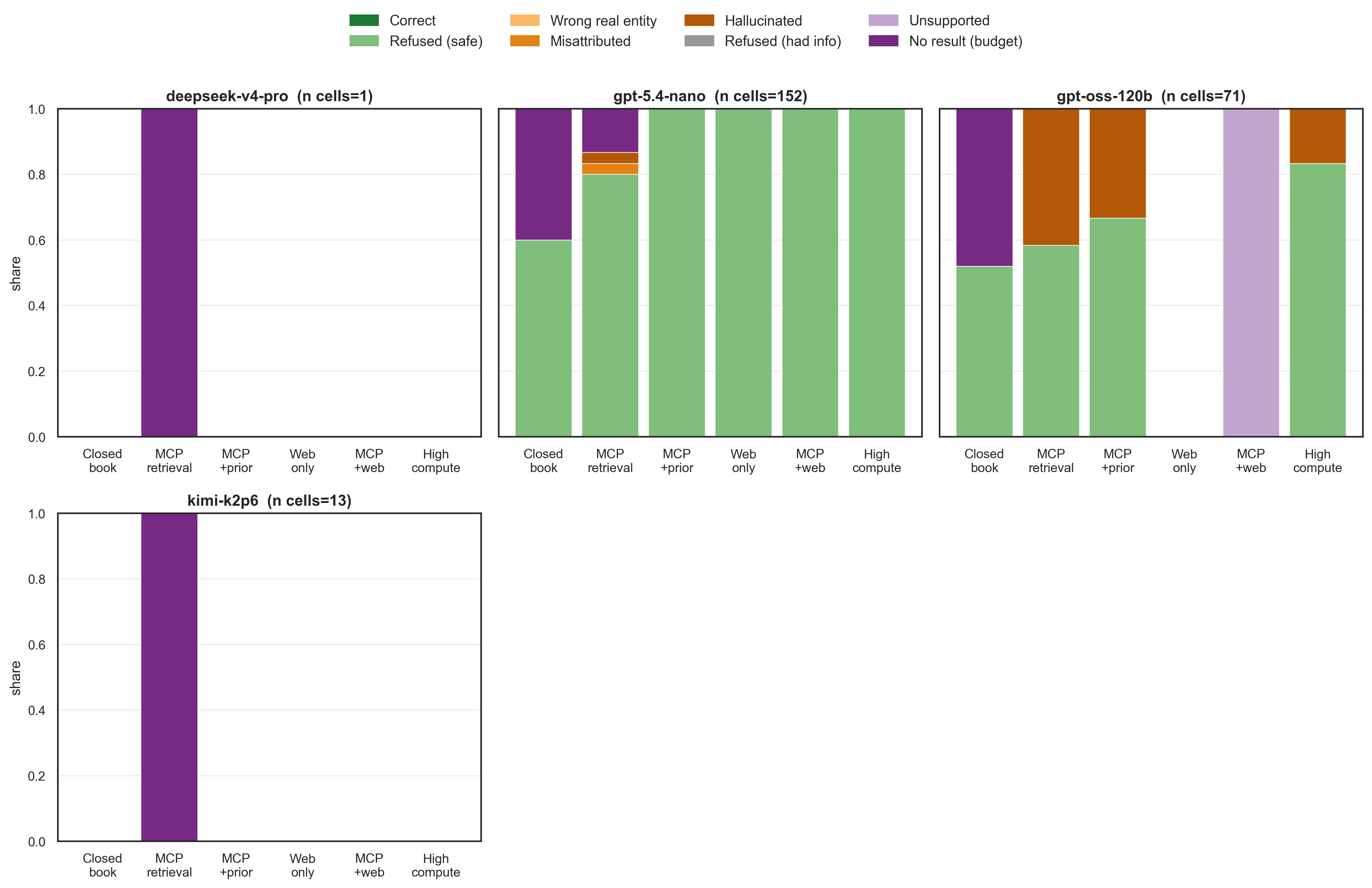}
\caption{\textbf{Tested rescue strategies did not recover hidden author links.} Author attribution with the People facet masked, shown separately for each model. The arms used memory-only answering with no tools, Model Context Protocol (MCP) retrieval, MCP plus prior, web only, MCP plus web, or a higher-compute MCP arm. No completed run returned the correct hidden author edge: 0 of 237 completed runs and 0 of 400 attempted runs were correct. Completion was uneven across models, so both denominators matter. Colors show whether a run was correct, refused safely, returned a wrong real entity, misattributed an edge, hallucinated, refused despite available information, was unsupported, or exhausted its budget.}
\label{fig:substitution}
\end{figure}

A natural objection is that metadata repair may be unnecessary if models can infer or search around missing metadata. We tested that objection directly for author attribution with the People facet hidden. The rescue arms included memory-only answering, Model Context Protocol (MCP) retrieval, MCP plus the model's own prior, web search only, MCP plus web search, and a higher-compute MCP arm with a 60-call budget (Fig.~\ref{fig:substitution}; Supplementary Methods, section~\ref{sec:appendix-substitution}).

The rescue arms changed how agents failed, not whether the hidden link came back. No completed run returned the correct hidden author edge: 0 of 237 completed runs and 0 of 400 attempted runs were correct. Some arms refused safely, some exhausted their budget, and some produced unsupported or hallucinated outputs, but none recovered the deliberately hidden author edge. The practical implication is not that compute can never substitute for metadata. It is that downstream systems should not assume inference-time search will reliably repair missing scholarly records on demand.

\begin{figure}[tbp]
\centering
\includegraphics[width=\linewidth]{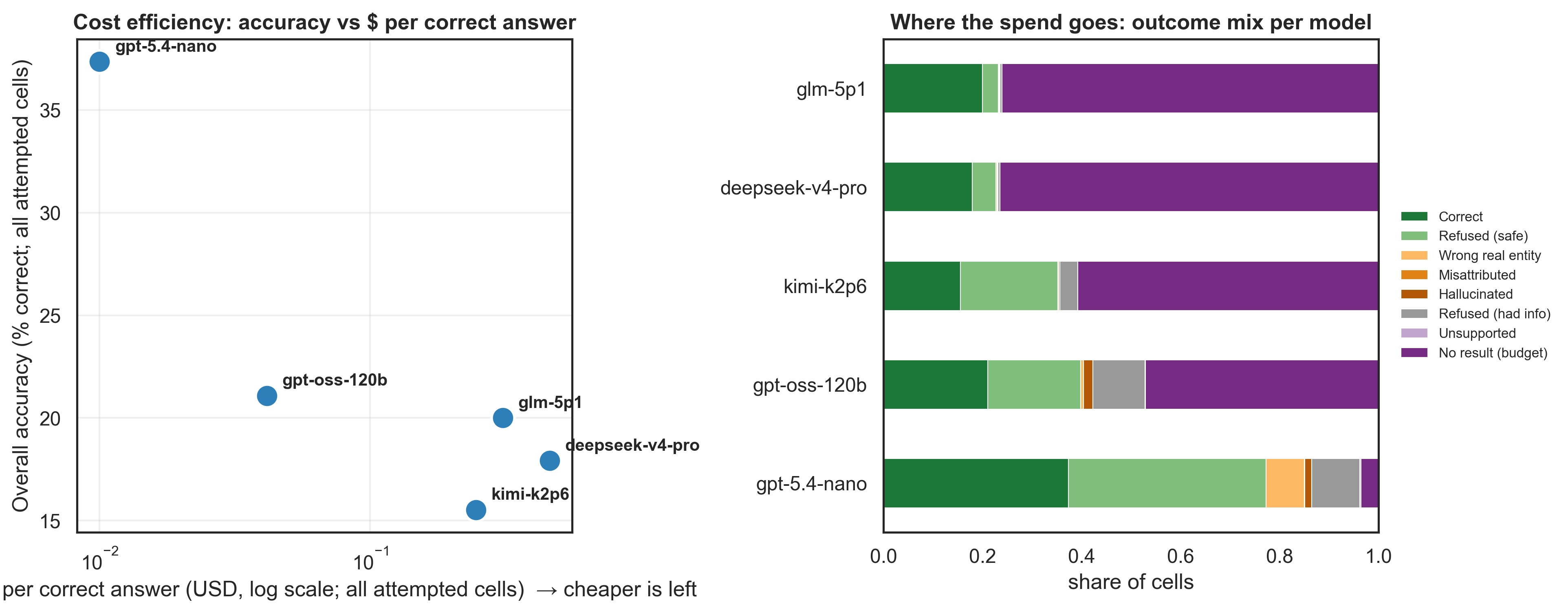}
\caption{\textbf{Additional inference cost did not compensate for missing scholarly links.} Left: overall accuracy over attempted runs versus dollars per correct answer. Right: outcome mix per model, including budget-exhausted runs that incurred cost. Cost per correct answer spanned roughly 46-fold when budget-exhausted runs were included, and much of the expensive model use ended without a result.}
\label{fig:cost}
\end{figure}

The cost result shows why this matters for deployment (Fig.~\ref{fig:cost}). Reasoning models and tool-heavy arms spent more tokens and dollars, but much of the additional work ended without a result. Per attempted cell, the reasoning and tool-heavy models were roughly 10- to 22-fold more expensive than the cheapest instruction-tuned baseline in this run. Across the tested systems, cost per correct answer spanned roughly 46-fold when budget-exhausted runs were included. This figure reports token and dollar cost, not environmental cost, which would require separate assumptions about energy use and hardware \cite{strubell2019energy}.

% =====================================================================
\section{Record verification points to repair}
\label{sec:discussion}

The experiments converge on one mechanism. A retrieved scholarly substrate constrained citation fabrication. Restoring a metadata facet restored the task that needed that facet. Removing the connection led to invented answers, refusals, or budget exhaustion, depending on the model and tools. Natural low-scoring OpenAlex records often lacked the same kinds of edges, so many attribution tests could not even be constructed from that substrate. Tested rescue strategies, including web search and extra tool calls, did not recover hidden author links. Thus the paper does not assume that record verification matters. It tests the premise: the record connections visible to an agent can determine what the agent can attribute.

This is the role proposed for \nexus. It is not a score of scientific merit, importance, venue prestige, citation impact, or author quality. It is a record-level check for whether a scholarly record exposes the connections needed for AI-mediated attribution, reading, and reuse. A paper can be scientifically important and still have a weak record. Conversely, a complete record does not make the science important. The measurement is narrower: can the agent see the author, affiliation, funder, reference, identifier, license, abstract, or text route it would need to support a particular scholarly claim?

This role places \nexus\ within the FAIR tradition. FAIR principles say that research objects should be findable, accessible, interoperable, and reusable; implementation profiles and community standards help fields build the workflows that make this possible. \nexus\ asks a narrower downstream question: after metadata enter public scholarly infrastructure, are the record connections needed for attribution, reading, and reuse actually visible to the agents that query them? A high score is not proof of full FAIR compliance. It is evidence that selected FAIR-relevant connections are present, resolvable, queryable, and traceable at scale. A low score points to places where FAIR intentions have not yet become usable structure in the scholarly graph.

That measurement points naturally to repair. If a publisher, archive, funder, or institution can see that its records lack Open Researcher and Contributor ID (ORCID) iDs, Research Organization Registry identifiers, funder identifiers, award numbers, reference lists, abstracts, licenses, or full-text links, those gaps can be repaired upstream. Repair still requires maintenance because identifiers can change and graphs can disagree. But a repaired record can serve many future uses. Inference-time reconstruction is attempted again for each query, and in the rescue setting tested here it failed while adding token and dollar cost.

The same logic extends beyond attribution, but the evidence is weaker there. Access is the reading version of the problem: a system can cite and read a full-text open record, cite shallowly from an abstract, or refuse when automated access is blocked. Our access experiment illustrates this direction but uses one model, a different protocol, and endogenous "touched DOI" denominators, so it is treated as a directional probe (Tables~\ref{tab:access-views} and \ref{tab:access-results}). Visibility is broader still. Science-of-science work has long studied how attention and cumulative advantage shape what is seen and cited \cite{merton1968matthew,fortunato2018sciofsci}. A small literature-review probe returned works that skewed toward higher Nexus-Score values, but it was underpowered and hypothesis-generating (Fig.~\ref{fig:matthew}). The present paper demonstrates a record-level attribution mechanism; it does not quantify population-scale visibility or equity effects.

The experiments also set boundaries for the next study. They motivate the need for a verifier, but they do not validate every choice in the final score. The composite \nexus\ value, the equal facet weights and prespecified signal weights in this OpenAlex implementation, and cross-substrate generality remain to be tested. The masking experiments require a leak-audited serving layer. Some reasoning models completed few runs under the 10-call cap. Natural sparsity results are OpenAlex task-buildability results, not audits of all possible sources. The access and visibility probes are extensions, not main causal evidence. The results also imply that attribution benchmarks should report the substrate they sample from, because ground-truth questions are easiest to build where records already contain the relevant links.

The practical lesson is that agents cannot reliably credit what the scholarly record does not expose. \nexus\ is proposed as an instrument for making those missing connections visible before they become downstream failures of attribution, reading, or reuse. The next work is to validate score bands against held-out outcomes, run the same design on other scholarly substrates, scale visibility experiments with field/language/region stratification, and connect measurement to metadata repair workflows.

% =====================================================================
\section*{Acknowledgments}
\textbf{Funding:} The authors are grateful to Lossfunk, OpenAI, Fireworks, and Y Combinator Startup School for providing API credits used to run the experiments.
\textbf{Author contributions:} Aadi~Narayana~Varma~Dantuluri led the conceptualization and experimental design, developed the software, conducted the majority of the experiments and data evaluation, and led manuscript drafting and revision. Sushrut~Thorat contributed to experimental design, conducted evaluation and validation, and contributed to manuscript drafting and revision. Paras~Chopra contributed to experimental design, interpretation of the results, and manuscript review and revision.
\textbf{Competing interests:} Aadi~Narayana~Varma~Dantuluri, the first and corresponding author, is the founder of the \nexus\ initiative (\href{https://nexus-score.org}{nexus-score.org}), which develops the measure evaluated in this paper. This constitutes a competing interest. To mitigate it, \nexus\ is defined transparently with its weights and operational signals pinned in a public configuration file, all run records and figure-generation code will be made available for independent verification, and the measure is deliberately blind to prestige, age, and citation count so that it cannot be tuned to favor particular works or venues. The remaining authors declare no competing interests.
\begin{placeholderbox}
\textbf{Data and materials availability:} Code, frozen configurations, corpus manifests, and analysis scripts are available at \href{https://github.com/aadivar/nexus-score-arxiv-preprint-code}{github.com/aadivar/nexus-score-arxiv-preprint-code}. The replication-package record is registered in Zenodo under the stable concept DOI \href{https://doi.org/10.5281/zenodo.21289889}{10.5281/zenodo.21289889}.
\end{placeholderbox}
\textbf{Additional acknowledgments:} The authors thank Alice Meadows (ORE Consulting) and Rob Johnson (Research Consulting, \href{https://research-consulting.com}{research-consulting.com}) for critical review of the draft, and Dhruv Trehan (Lossfunk) and Pranay Kundu (Founder, Yudai Labs, \href{https://www.yudai.app}{yudai.app}) for their input. The authors also thank the open scholarly infrastructure community, including contributors to OpenAlex, Crossref, ROR, ORCID, DOAJ, and the broader open research information ecosystem.

% =====================================================================

% =====================================================================
\clearpage
\appendix
\begin{center}
{\large\bfseries\color{accent} Supplementary Information}
\end{center}
\vspace{0.6em}

\section{Experimental setup in detail}
\label{sec:appendix-setup}

This supplement collects the technical details needed to reproduce or audit the main claims. It first gives the common setup used across experiments, then gives result-specific protocols in the same order as the main text, then lists supplementary result tables, scoring details, and reproducibility files.

We use \emph{run} or \emph{cell} for one model applied to one work, task, view, and experimental arm. An \emph{arm} is a tested strategy, such as retrieval-only, web search, or higher-compute retrieval. Bracketed intervals are 95\% Wilson score intervals for binomial proportions. File paths are included so that each reported number can be traced to its intended source file when the data and code package is finalized.

\subsection{Models and decoding}
\label{sec:appendix-models}

The main attribution, diagonal, substitution, and cost experiments use five model families: \code{gpt-5.4-nano}, \code{gpt-oss-120b}, DeepSeek~V4~Pro (\code{deepseek-v4-pro}), Kimi~K2.6 (\code{kimi-k2p6}), and GLM-5.1 (\code{glm-5p1}). For these runs the client decodes at temperature $0.0$ with at most $4{,}096$ output tokens per turn and a 120-second timeout. Temperature controls how variable a model's next-token choices are; here, the main experiments use the lowest-temperature setting available in the runner. The no-retrieval contrast and access probe use \code{gpt-5.5-2026-04-23}, decoded at temperature $1.0$ with \code{max\_completion\_tokens = 10{,}240}. Some rescue arms additionally enable live web search through the Parallel AI Search API \cite{parallel_search}, so lack of web access is not the explanation for failure in those arms.

\begin{table}[H]
\centering
\small
\begin{tabularx}{\linewidth}{Xllrrr}
\toprule
Model (\code{name}) & Provider & Model weights & Context (k) & \$/1M in & \$/1M out \\
\midrule
\code{deepseek-v4-pro} & Fireworks & open & 131 & 1.74 & 3.48 \\
\code{gpt-oss-120b} & Fireworks & open & 131 & 0.15 & 0.60 \\
\code{kimi-k2p6} & Fireworks & open & 131 & 0.95 & 4.00 \\
\code{glm-5p1} & Fireworks & open & 131 & 1.40 & 4.40 \\
\code{gpt-5.4-nano} & OpenAI & closed & 200 & 0.20 & 1.25 \\
\midrule
\code{gpt-5.5-2026-04-23}$^{\dagger}$ & OpenAI & closed & 200 & 2.50 & 15.00 \\
\bottomrule
\end{tabularx}
\caption{Frozen model slate and prices captured on 2026-05-24. ``Open'' means the model weights are publicly available; ``closed'' means they are not. Context is the maximum input/output window, in thousands of tokens. $^{\dagger}$\code{gpt-5.5-2026-04-23} is used only for the no-retrieval and access probes. Prices are used for the cost analysis; exact model identifiers and decoding parameters are logged with every run.}
\label{tab:repro-models}
\end{table}

\subsection{Interface, tools, and tool budgets}
\label{sec:appendix-interface}

The agent reaches the substrate through a Model Context Protocol (MCP) shim between the model and OpenAlex \cite{openalex2022}. A shim is a small serving layer that sits between the model and the underlying database. Here it serves real records but can hide selected fields on a per-view basis, which is how masking and restoration are implemented on identical works. The attribution tools expose search, entity fetch (\code{get\_work}, \code{get\_author}), and relation lookups. In every arm the model is driven through function calling, meaning that the model requests a tool call and receives the tool result before continuing. The models with public weights are served through Fireworks and the GPT models through OpenAI directly.

A \emph{tool call} is one query whose result is fed back to the model. The standard cap is 10 calls per question for the MCP retrieval, retrieval-plus-prior, and web-only arms. The combined MCP-plus-web arm allows 20 calls. The access experiment allows 30 calls. The higher-compute substitution arm raises the cap to 60 calls. \code{NO\_RESULT} denotes a run in which the agent exhausted its tool budget without returning a final answer.

The access experiment uses a separate tool loop. Its \code{fetch\_url} tool follows OpenAlex open-access pointers and retrieves full text from PubMed Central open-access records through the NCBI E-utilities \code{efetch} API \cite{ncbi_eutilities}, with abstracts reconstructed from the abstract text stored in OpenAlex where needed. An open-access pointer is a machine-readable route from the record to a version of the paper that can be reached without a paywall.

\subsection{Views: masking and restoration}
\label{sec:appendix-views}

A \emph{view} is the metadata actually served for a work. The attribution experiments use two view families over the same works. Masked views start from the full record and hide one facet: \code{V\_full}, \code{V\_people\_masked}, \code{V\_organizations\_masked}, and \code{V\_funding\_masked}. Restored views start from a minimal record and add one facet back: \code{V\_minimal}, \code{V\_minimal\_plus\_people}, \code{V\_minimal\_plus\_organizations}, \code{V\_minimal\_plus\_funding}, and \code{V\_minimal\_plus\_provenance}. The six controlled access views are listed in Supplementary Table~\ref{tab:access-views}.

Controlled masking only works if the serving layer is consistent. The shim masks entity records consistently with the work view: \code{get\_author} strips \code{last\_known\_institutions} under People-only views, \code{raw\_affiliation\_strings} are stripped wherever Organizations is not exposed, and \code{verify\_work\_entity\_edge} refuses relation probes of non-exposed facets. Apparent mismatched-facet successes were audited down to the tool trace before interpreting the diagonal. One boundary case is retained as a design lesson: in an Organizations-masked view, one Kimi run recovered an institution because the work was deposited in an institutional archive whose source record exposed \code{host\_organization}. That is not a counting error; it shows that real metadata graphs can carry the same institutional fact through more than one facet.

\subsection{Tasks and outcomes}
\label{sec:appendix-tasks}

Author attribution, institution attribution, and funding attribution ask the agent to name the author, institution, or funder connection for a work. Citation lineage asks it to list works referenced by a target work. Known-item DOI lookup asks it to recover the DOI of a specifically named work and serves as a negative control. Two additional tasks support the directional extensions: the literature-review task for visibility and the access task, in which the agent files every touched paper as read, abstract-only, unread, inaccessible, or unused. In the access task, the response contract required \code{GOOD}, \code{BAD}, and \code{UGLY} sections: papers could be cited only after a successful fetch, abstract-only uses had to be labeled, and unreachable papers had to be listed rather than silently cited. A cited DOI without a successful fetch was flagged by the adjudication script as cited but unread.

\begin{table}[H]
\centering
\small
\begin{tabularx}{\linewidth}{lX}
\toprule
Bucket & Meaning \\
\midrule
\code{CORRECT} & The asserted connection holds in the ground-truth graph. \\
\code{REFUSED\_CORRECTLY} & The agent declined because the needed metadata was hidden or absent. \\
\code{REFUSED\_INCORRECTLY} & The agent declined although the needed information was available. \\
\code{WRONG\_REAL} & A real but incorrect entity was returned. \\
\code{MISATTRIBUTED} & A connection was asserted between real entities but does not hold. \\
\code{HALLUCINATED} & A non-existent entity or identifier was invented. \\
\code{UNSUPPORTED} & The answer was not backed by a fetched or queried record. \\
\code{NO\_RESULT} & The tool budget was exhausted before a final answer. \\
\bottomrule
\end{tabularx}
\caption{Outcome buckets. The decisive test is connection validity: a real author and a real work are not enough unless the relation between them holds in the ground-truth graph.}
\label{tab:appendix-outcomes}
\end{table}

The detailed outcome taxonomy in Fig.~\ref{fig:outcome} is reported for \code{gpt-5.4-nano}, the strongest grounded agent in the tested loop. Cross-model failure information comes from the restoration table, the substitution experiment, and the cost/outcome panel. Those results show that missing links do not produce one universal behavior. Depending on the model and tool setting, the agent may invent, refuse, or stop without a final answer. The common point is narrower and stronger: when the required record connection is unavailable, the requested attribution is not recovered.

\subsection{Record pools}
\label{sec:appendix-pools}

The gold pool contains 142 richly described works (mean Nexus-Score composite approximately 0.95) where every tested connection exists before masking. The natural low-scoring pool contains 175 works, 25 per domain, drawn from the open record as-is and filtered only to require a DOI; a 25-work evaluated subset drives the buildability result. The p53 corpus is the retrieval candidate list for the no-retrieval contrast. Sampling is seeded with \code{20260524}; all records are drawn from OpenAlex snapshot \code{live-2026-05-24}.

\subsection{No-retrieval grounding protocol}
\label{sec:appendix-norag}

The no-retrieval contrast in Section~\ref{sec:norag} is a call-level citation check, not a balanced benchmark. The same model, \code{gpt-5.5-2026-04-23}, answered handwritten questions about p53 and apoptosis under two conditions. In the retrieval condition, each call received a candidate list of real papers from the p53 corpus and was instructed to cite only papers from that list. In the no-retrieval condition, the model received no candidate list and had to cite from memory. Each answer had to provide DOIs, so a script could parse the cited identifiers, compare them with the p53 corpus, and test out-of-corpus identifiers against \code{doi.org}.

The retrieval arm ran ten repetitions of ten questions, for 100 calls. The no-retrieval arm was stopped after ten repetitions of the first two questions, for 20 calls. Both arms used temperature $1.0$ and \code{max\_completion\_tokens = 10{,}240}. Cited DOIs were parsed from the model outputs. A DOI counted as \emph{in corpus} only if it matched a DOI in the p53 candidate corpus. A DOI outside the corpus was checked with a \code{HEAD} request to \code{doi.org}: resolving DOIs were counted as real identifiers outside the corpus, whereas non-resolving DOIs counted toward the fabricated-reference rate. Thus the ``doi.org resolve, not in corpus'' bucket means real but unverified against the task corpus, not confirmed relevant to the question.

\subsection{Substitution arms}
\label{sec:appendix-substitution}

The substitution test runs six rescue strategies on author attribution with the People facet hidden: memory-only answering with no tools, MCP retrieval, MCP plus the model's own prior, web search only, MCP plus web search, and a higher-compute MCP arm with the 60-call budget. The two web arms reach the open web through a \code{web\_search} tool backed by the Parallel AI Search API endpoint \code{/v1beta/search}, using \code{fast} mode and up to 10 results per call \cite{parallel_search}. Provider settings are pinned in \code{config/web\_search.yaml} and the key is read from \code{PARALLEL\_API\_KEY}. Per-request web-search costs are folded into each run's dollar total.

Across the 400 attempted substitution cells, 237 completed with a final answer and none returned the hidden author edge. Completed cells were unevenly distributed across models: 152 for \code{gpt-5.4-nano}, 71 for \code{gpt-oss-120b}, 13 for \code{kimi-k2p6}, 1 for \code{deepseek-v4-pro}, and 0 for \code{glm-5p1}. The Wilson upper bound for the correct-rescue rate is 1.6\% on completed cells and 0.95\% on attempted cells.

\section{Supplementary result tables}

\begin{figure}[H]
\centering
\includegraphics[width=\linewidth]{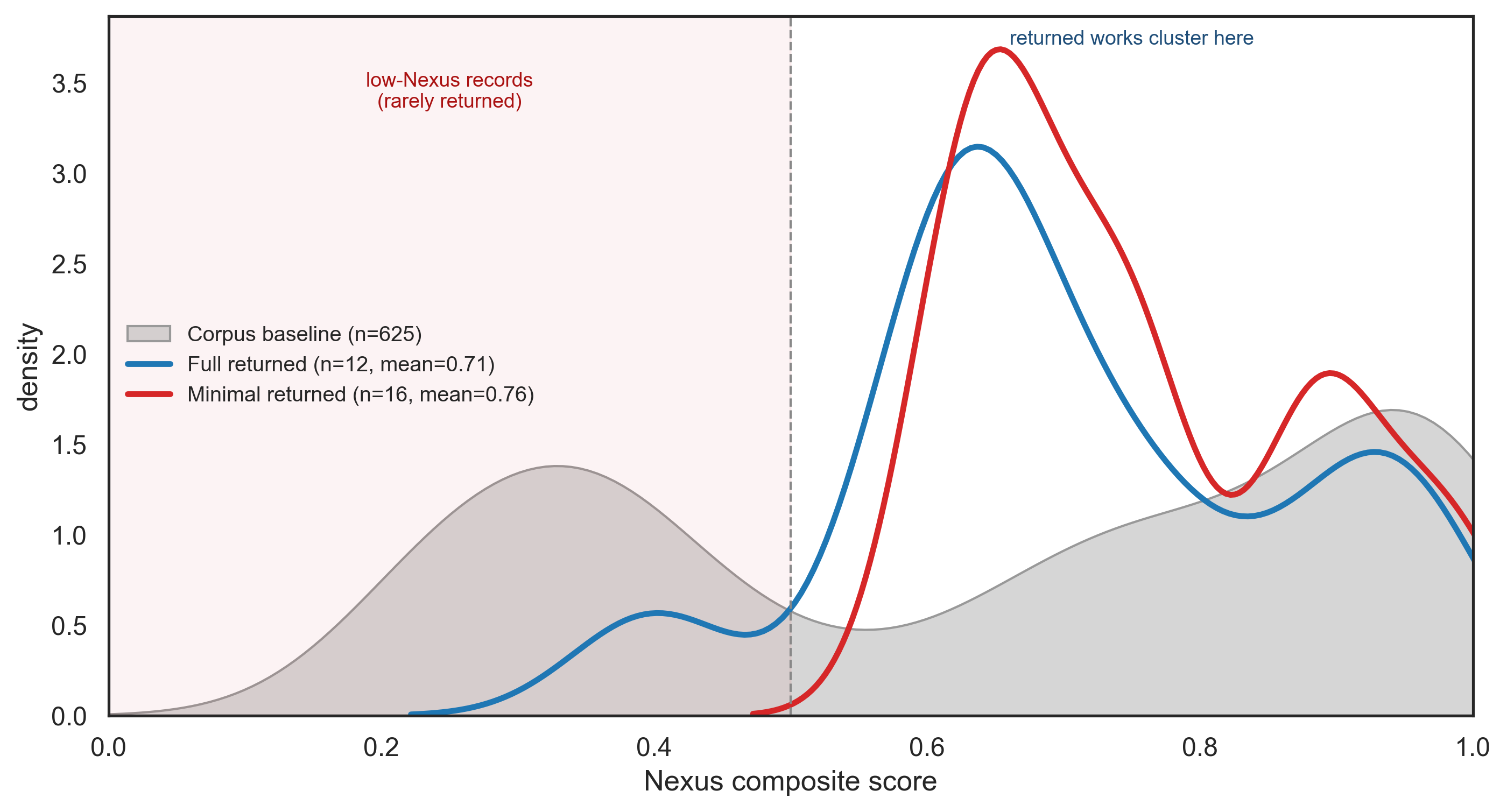}
\caption{\textbf{Exploratory visibility probe.} Returned works in a small literature-review task skewed toward higher Nexus-Score values than the full corpus distribution: 1 of 28 scored returned works fell below 0.5, whereas 38\% of the corpus did. The full-record view returned 12 scored works with mean Nexus-Score 0.71; the minimal-record view returned 16 scored works with mean 0.76. The sample is small, and the agent returned about two works on average despite being asked for ten. This figure is hypothesis-generating and does not establish a population-scale Matthew effect.}
\label{fig:matthew}
\end{figure}

This section follows the result order of the main text. Supplementary Table~\ref{tab:norag} gives the full retrieval-grounding contrast. Supplementary Tables~\ref{tab:diag} and \ref{tab:citation} give denominators and intervals for facet restoration and citation lineage. Supplementary Tables~\ref{tab:access-views} and \ref{tab:access-results} give the access-probe views and outcomes discussed in the final section. Supplementary Fig.~\ref{fig:matthew} gives the exploratory visibility probe. Supplementary Table~\ref{tab:appendix-substrate} gives the corpus-quality distribution used to motivate \nexus\ as a verifier.

\begin{table}[H]
\centering
\small
\begin{tabularx}{\linewidth}{lrrrrr}
\toprule
Arm & Calls & \makecell[r]{Cited DOIs\\(total / unique)} & In corpus & \makecell[r]{doi.org resolve\\(not in corpus)} & \makecell[r]{Fabricated\\($\geq1$ per call)} \\
\midrule
Retrieval context & 100 & 581 / 70 & \textcolor{pos}{\textbf{70 (100\%)}} & 0 & \textcolor{pos}{\textbf{0 / 100 (0\%)}} \\
No retrieval (early-stopped) & 20 & 283 / 65 & \textcolor{neg}{0 (0\%)} & 43 (66.2\%) & \textcolor{neg}{\textbf{15 / 20 (75\%)}} \\
\bottomrule
\end{tabularx}
\caption{Retrieval-constrained citation contrast. The retrieval arm covered ten repetitions of ten questions; the no-retrieval arm was stopped after ten repetitions of the first two questions, so this is a motivating contrast rather than a balanced comparison. With a candidate record, every cited DOI was in the corpus and no call contained a fabricated DOI (0\%, 95\% Wilson interval $[0, 3.7]$\%). Without retrieval, none of the 65 unique cited DOIs was in the corpus, and 15 of 20 calls contained at least one fabricated DOI (75\%, $[53, 89]$\%).}
\label{tab:norag}
\end{table}

\begin{table}[H]
\centering
\small
\begin{tabularx}{\linewidth}{lrrX}
\toprule
Model & \makecell[r]{Matched\\(diagonal)} & \makecell[r]{Mismatched} & Notes \\
\midrule
\code{gpt-5.4-nano} & \textcolor{pos}{\textbf{97\%}} \scriptsize[91, 99] & \textcolor{pos}{\textbf{0\%}} \scriptsize[0, 2] & cleanest pattern \\
\code{deepseek-v4-pro} & 92\% \scriptsize[65, 99] & 0\% \scriptsize[0, 12] & few completed cells \\
\code{glm-5p1} & 87\% \scriptsize[62, 96] & 0\% \scriptsize[0, 11] & few completed cells \\
\code{gpt-oss-120b} & 62\% \scriptsize[52, 72] & 0\% \scriptsize[0, 4] & moderate signal \\
\code{kimi-k2p6} & 22\% \scriptsize[13, 34] & 0\% \scriptsize[0, 3] & high \code{NO\_RESULT} \\
\bottomrule
\end{tabularx}
\caption{Matched versus mismatched restoration over completed cells. Matched/mismatched completed-cell denominators are $115/180$ (\code{gpt-5.4-nano}), $12/29$ (\code{deepseek-v4-pro}), $15/31$ (\code{glm-5p1}), $80/99$ (\code{gpt-oss-120b}), and $55/130$ (\code{kimi-k2p6}). Attempted matched/mismatched denominators were $115/180$ for \code{gpt-5.4-nano} and \code{gpt-oss-120b}, and $90/180$ for \code{deepseek-v4-pro}, \code{glm-5p1}, and \code{kimi-k2p6}; figure-panel averages can therefore differ slightly from pooled completed-cell values. Across all five models, zero of 469 completed mismatched cells was correct.}
\label{tab:diag}
\end{table}

\begin{table}[H]
\centering
\small
\begin{tabularx}{\linewidth}{lrrr}
\toprule
View & \code{gpt-5.4-nano} & \code{gpt-oss-120b} & Cells \\
\midrule
\code{V\_full} & 92\% \scriptsize[75, 98] & 64\% \scriptsize[45, 80] & 50 \\
\code{V\_citation\_masked} & 0\% \scriptsize[0, 13] & 0\% \scriptsize[0, 13] & 50 \\
\code{V\_minimal} & 0\% \scriptsize[0, 13] & 0\% \scriptsize[0, 13] & 50 \\
\code{V\_minimal\_plus\_provenance} & 96\% \scriptsize[80, 99] & 100\% \scriptsize[87, 100] & 50 \\
\bottomrule
\end{tabularx}
\caption{Citation lineage, \% \code{CORRECT} per model and view with 95\% Wilson intervals. Restoring Provenance to a minimal record restores the task; hiding it removes the reference connection needed for the task.}
\label{tab:citation}
\end{table}

\begin{table}[H]
\centering
\small
\begin{tabularx}{\linewidth}{lcccX}
\toprule
View & Abstract? & Open-access URL? & Fetch result & Real-world analog \\
\midrule
\code{V\_full\_access} & yes & yes & \code{FULL\_TEXT} & machine-readable open access \\
\code{V\_abstract\_only} & yes & no & \code{NO\_URL} & indexed but no resolvable open-access URL \\
\code{V\_no\_abstract} & no & yes & \code{FULL\_TEXT} & missing abstract, working open-access URL \\
\code{V\_paywalled} & yes & yes & \code{ABSTRACT\_ONLY} & closed article, abstract accessible \\
\code{V\_robots\_deny} & yes & yes & \code{FORBIDDEN\_403} & automated access blocked \\
\code{V\_metadata\_only} & no & no & \code{NO\_URL} & title and DOI only \\
\bottomrule
\end{tabularx}
\caption{Controlled access views used in the directional access probe.}
\label{tab:access-views}
\end{table}

\begin{table}[H]
\centering
\small
\begin{tabularx}{\linewidth}{lrrrrrr}
\toprule
View & \makecell[r]{Touched\\DOIs ($n$)} & \makecell[r]{Cited\\and read} & \makecell[r]{Cited from\\abstract} & \makecell[r]{Cited unread} & \makecell[r]{Refused\\no access} & \makecell[r]{Found but\\unused} \\
\midrule
\code{V\_full\_access} & 110 & 31.8\% & 0\% & 6.4\% \scriptsize[3.1, 12.6] & 52.7\% & 9.1\% \\
\code{V\_abstract\_only} & 142 & 0\% & 0\% & 5.6\% \scriptsize[2.9, 10.7] & 88.0\% & 6.3\% \\
\code{V\_no\_abstract} & 111 & 27.9\% & 0\% & 7.2\% \scriptsize[3.7, 13.6] & 53.2\% & 11.7\% \\
\code{V\_paywalled} & 82 & 0\% & 70.7\% & 1.2\% \scriptsize[0.2, 6.6] & 15.9\% & 12.2\% \\
\code{V\_robots\_deny} & 84 & 0\% & 0\% & 7.1\% \scriptsize[3.3, 14.7] & 92.9\% & 0\% \\
\code{V\_metadata\_only} & 76 & 0\% & 0\% & 14.5\% \scriptsize[8.3, 24.1] & 84.2\% & 1.3\% \\
\bottomrule
\end{tabularx}
\caption{Access-probe outcomes over touched DOIs. The denominator is endogenous because the agent touched different DOI sets under different views; this table is therefore directional, not a fixed-DOI causal estimate. Brackets give 95\% Wilson intervals for the cited-unread column.}
\label{tab:access-results}
\end{table}

The original exploratory notes also record an anecdotal observation that several surfaced works appeared to have high online attention outside the citation graph. We do not quantify that observation here and make no claim from it. A larger visibility study should model agent visibility against Nexus-Score, citation count, and external attention measures jointly, so metadata-mediated visibility can be separated from ordinary public-attention effects.

\section{Nexus scoring and corpus quality}
\label{sec:appendix-substrate}

The main text uses Nexus-Score to ask whether a record exposes the connections needed for attribution, reading, and reuse. The experimental Nexus-Score composite is an equal-weighted average of five facet scores; each facet is a prespecified weighted average of five signal values in $[0,1]$:
\[
  \nexus(r)=\sum_{f} w_f F_f(r),
  \qquad
  F_f(r)=\sum_s v_{f,s}\sigma_{f,s}(r),
\]
with $w_f=\tfrac{1}{5}$ for every facet and $(v_{\mathrm{presence}},v_{\mathrm{resolvability}},v_{\mathrm{specificity}},v_{\mathrm{queryability}},v_{\mathrm{provenance}})=(0.30,0.25,0.20,0.15,0.10)$ within every facet. This differs from the public Crossref/publisher-level Nexus-Score implementation, which uses publisher-level coverage statistics and nonuniform dimension weights: Provenance 25\%, People 20\%, Organizations 15\%, Funding 20\%, and Access 20\%. The present experiment tests the record-level premise behind the broader verifier rather than the public-site scoring formula.

\begin{table}[H]
\centering
\small
\begin{tabularx}{0.86\linewidth}{Xrr}
\toprule
Facet & Mean & Median \\
\midrule
Provenance & 0.91 & 1.00 \\
People & 0.72 & 0.94 \\
Organizations & 0.60 & 1.00 \\
Access & 0.69 & 0.86 \\
Funding & 0.34 & 0.00 \\
\midrule
\textbf{Composite} & \textbf{0.65} & \textbf{0.73} \\
\bottomrule
\end{tabularx}
\caption{Substrate quality of the working corpus ($n=625$). The composite has standard deviation $0.28$, interquartile range $0.37$--$0.92$, and full range $0.20$--$1.00$. By band, $14.1\%$ of works score below $0.30$, $24.0\%$ fall in $[0.30,0.50)$, $22.9\%$ in $[0.50,0.80)$, and $39.0\%$ at $0.80$ or above.}
\label{tab:appendix-substrate}
\end{table}

\section{Reproducibility}
\label{sec:repro}

Everything that affects a result is specified in the files listed below, and the per-run JSON record is the source artifact for each run. Corpus tables, summary parquet files, and figures are derived from those records. Parquet is a column-oriented table format used here for derived data tables.

\subsection{Prerequisites}

The analysis requires Python $\geq 3.12$, \code{uv}, and network access to the OpenAlex API. Corpus construction requires \code{OPENALEX\_MAILTO}; an \code{OPENALEX\_API\_KEY} is optional. The agent-run phase additionally requires \code{FIREWORKS\_API\_KEY}, \code{OPENAI\_API\_KEY}, and \code{PARALLEL\_API\_KEY}. \code{ANTHROPIC\_API\_KEY} and \code{GEMINI\_API\_KEY} are wired for extension but are not required for the reported runs. Credentials are read from a local \code{.env} file and are not included in the data and code package.

Dependencies are declared in \code{pyproject.toml} and locked in \code{uv.lock}. The core install pins the corpus and scoring stack; the \code{agents} extra adds model SDKs; and the \code{analysis} extra adds plotting and statistical packages. After \code{.env} is configured, \code{scripts/verify\_keys.py} checks provider connectivity without printing key values.

\subsection{Frozen inputs}

The substrate is the OpenAlex snapshot \code{live-2026-05-24}. Sampling is seeded with \code{20260524}, the inclusion window is publication years 2015--2024, and the pilot draws 25 works per pool per domain. The full set of frozen inputs is:

\begin{table}[H]
\centering
\small
\begin{tabularx}{\linewidth}{lX}
\toprule
Input & Location \\
\midrule
Snapshot date, seed, pool sizes, inclusion window & \code{config/preregistration.yaml} \\
Nexus-Score facet weights and signal definitions & \code{config/nexus\_weights.yaml} \\
Model identifiers, decoding, and cost parameters & \code{config/models.yaml} \\
Domain-to-topic mappings & \code{config/domains.yaml} \\
Web-search provider settings & \code{config/web\_search.yaml} \\
Selected work identifiers per pool & \code{data/corpus/v1/*.parquet} \\
Corpus build provenance & \code{data/corpus/v1/manifest.json} \\
Per-run records & \code{data/runs/v1/.../*.json} \\
\bottomrule
\end{tabularx}
\caption{Frozen inputs used to build the corpus, compute Nexus-Score values, run the agent cells, and regenerate figures. Raw OpenAlex records and per-run caches are rebuilt from the manifest and are therefore not the source result records.}
\label{tab:repro-frozen}
\end{table}

\subsection{End-to-end commands}

\begin{lstlisting}[language=bash]
# 0. Environment and credentials, run inside the project directory
uv sync --extra agents --extra analysis
cp .env.example .env
uv run python scripts/verify_keys.py

# 1. Build the substrate, scores, and study pools
uv run nexus snapshot resolve-topics
uv run nexus snapshot fetch
uv run nexus score
uv run nexus corpus build

# 2. Run experimental cells (example: facet restoration)
uv run nexus run list-tasks
uv run nexus run list-views
uv run nexus run pilot \
  --pool gold_edge --limit 30 \
  --task author_attribution --task institution_attribution \
  --task funding_attribution --task citation_lineage --task known_item \
  --view V_minimal --view V_minimal_plus_people \
  --view V_minimal_plus_organizations --view V_minimal_plus_funding \
  --view V_minimal_plus_provenance \
  --model gpt-5.4-nano --model gpt-oss-120b --model kimi-k2p6 \
  --model deepseek-v4-pro --model glm-5p1 \
  --arm B_mcp_rag --concurrency 10

# 3. Rebuild summary and figures
uv run python scripts/rebuild_summary.py
uv run python scripts/figures.py
\end{lstlisting}

The pilot command writes one JSON record per run under \code{data/runs/v1/...}. The substitution and visibility experiments use the same runner with different \code{--task}, \code{--view}, \code{--arm}, and \code{--model} flags. Because each JSON record is the source artifact for a run, \code{scripts/rebuild\_summary.py} can reconstruct \code{summary.parquet} from disk, and \code{scripts/figures.py} regenerates figures from the parquet summaries and the Nexus-Score table.

The no-retrieval and access probes run as standalone scripts:

\begin{lstlisting}[language=bash]
uv run python data/no_rag_hallucination/prompts/rerun_grounding_no_rag.py
uv run python data/no_rag_hallucination/prompts/resolve_cited_dois.py

uv run python data/access_experiment/prompts/access_experiment.py
uv run python data/access_experiment/prompts/adjudicate.py
\end{lstlisting}

Both standalone scripts read \code{OPENAI\_API\_KEY} and \code{OPENALEX\_MAILTO} from \code{.env}. The no-retrieval DOI resolver checks cited DOIs with requests to \code{doi.org}. The access runner checkpoints every five cells and caches OpenAlex queries and fetched PubMed Central bodies on disk, so re-execution is incremental. Its adjudication script buckets every touched DOI using rule-based checks with no language-model judge.

Language-model calls are not bit-reproducible, and hosted models may change over time. For this reason the exact model id is logged with every run, and \code{NO\_RESULT} is reported rather than dropped. The 60-cell access run consumed about $3.6$M input and $0.13$M output tokens over $1{,}061$ tool calls, took roughly 52 minutes, and cost approximately \$$10.9$ at the listed model prices; this cost is reported for transparency and is not part of the five-model cost comparison.

The cost analyses report tokens and dollars, not carbon emissions. Estimating emissions would require provider-specific information about hardware, utilization, datacenter energy mix, and power-usage effectiveness. Prior work shows that large neural-network training and inference can have material energy costs \cite{strubell2019energy}; the present study therefore treats environmental cost as a plausible implication of repeated inference-time repair, not as a measured outcome.

\end{document}